\documentclass[12pt]{article}
\usepackage{graphicx}
\usepackage{amsmath}
\usepackage{amssymb}
\usepackage{wrapfig}
\usepackage{sidecap}

\newcommand\pubnumber{SNSN-323-63}
\newcommand\pubdate{\today}

\def\napoli{$^1$NASA Postdoctoral Program Fellow, Goddard Space Flight Center, Greenbelt, MD 20771, USA\\
$^2$Department of Physics and Mathematics, College of Science and Engineering, Aoyama Gakuin University, 5-10-1 Fuchinobe, Chuo-ku, Sagamihara-shi, Kanagawa 252-5258, Japan\\
$^3$NASA Goddard Space Flight Center, Greenbelt, MD 20771, USA\\
$^4$Los Alamos National Laboratory, B244, Los Alamos, NM, 87545, USA\\
$^5$Department of Astronomy and Astrophysics, University of Chicago, Chicago, IL 60637, USA\\
$^6$Center for Research and Exploration in Space Science and Technology (CRESST) and NASA Goddard Space Flight Center, Greenbelt, MD 20771, USA\\
}
\def\support{\footnote{Work supported by an appointment to the NASA Postdoctoral Program at 
the Goddard Space Flight Center, administered by Oak 
Ridge Associated Universities through a contract with NASA.}}

\def\Title#1{\begin{center} {\Large #1 } \end{center}}
\def\Author#1{\begin{center}{ \sc #1} \end{center}}
\def\Address#1{\begin{center}{ \it #1} \end{center}}

\newcommand\pubblock{\rightline{\begin{tabular}{l} \pubnumber\\
         \pubdate  \end{tabular}}}
\newenvironment{Abstract}{\begin{quotation}  }{\end{quotation}}
\newenvironment{Presented}{\begin{quotation} \begin{center} 
             PRESENTED AT\end{center}\bigskip 
      \begin{center}\begin{large}}{\end{large}\end{center} \end{quotation}}
\def\Acknowledgements{\bigskip  \bigskip \begin{center} \begin{large}
             \bf ACKNOWLEDGEMENTS \end{large}\end{center}}

\def\beq{\begin{equation}}
\def\eeq#1{\label{#1}\end{equation}}
\def\eeqn{\end{equation}}

\def\beqa{\begin{eqnarray}}
\def\eeqa#1{\label{#1}\end{eqnarray}}
\def\eeqan{\end{eqnarray}}

\let\bar=\overbar

\def\Dslash{\not{\hbox{\kern-4pt $D$}}}
\def\dslash{\not{\hbox{\kern-2pt $\del$}}}

\def\msb{{\bar{\ssstyle M \kern -1pt S}}}

\begin{document}
\begin{titlepage}
\pubblock

\vfill
\Title{Probing the Gamma-Ray Burst Rate with Trigger Simulations of the {\it Swift} Burst Alert Telescope}
\vfill
\Author{Amy Lien\support,Takanori Sakamoto$^2$,Neil Gehrels$^3$,David M. Palmer$^4$,\\Scott~D.~Barthelmy$^3$,Carlo Graziani$^5$,John~K.~Cannizzo$^6$}
\Address{\napoli}
\vfill
\begin{Abstract}
The long gamma-ray burst (GRB) rate is essential for revealing the connection between GRBs, supernovae and stellar evolution. Additionally, the GRB rate at high redshift provides a strong probe of star formation history in the early universe. While hundreds of GRBs are observed by {\it Swift}, it remains difficult to determine the intrinsic GRB rate due to the complex trigger algorithm of {\it Swift}. Current studies usually approximate the {\it Swift} trigger algorithm by a single detection threshold. However, unlike the previously flown GRB instruments, {\it Swift} has over 500 trigger criteria based on photon count rate and additional image threshold for localization. To investigate possible systematic biases and explore the intrinsic GRB properties, we developed a program that is capable of simulating all the rate trigger criteria and mimicking the image trigger threshold. We use this program to search for the intrinsic GRB rate. Our simulations show that adopting the complex trigger algorithm of {\it Swift} increases the detection rate of dim bursts. As a result, we find that either the GRB rate is much higher than previously expected at large redshift, or the luminosity evolution is non-negligible. We will discuss the best results of the GRB rate in our search, and their impact on the star-formation history. 
\end{Abstract}
\vfill
\begin{Presented}
Huntsville Gamma-Ray Burst Symposium,\\
Nashville, Tennessee,  USA, April 14--18, 2013
\end{Presented}
\vfill
\end{titlepage}
\def\thefootnote{\fnsymbol{footnote}}
\setcounter{footnote}{0}

\section{Introduction}

The extraordinary luminosities of gamma-ray bursts (GRBs) and their connection with the deaths of massive stars (e.g.,~\cite{Galama98, Heger03})
make long GRBs 
unique and independent probes of the cosmic star-formation rate (SFR), especially at large redshifts
where it becomes difficult for other methods.
Many efforts have been done to map out the intrinsic cosmic GRB rate as a function of redshift from 
current observations (e.g., \cite{Yuksel08,Butler10,Wanderman10}).
Results from these studies suggest that the cosmic GRB rate generally follows the shape of the cosmic SFR
at low redshift. However, at large redshift ($z \gtrsim 4$), several groups have suggested 
a higher GRB rate than previous expectation based on former measurements of the SFR \cite{Yuksel08,Butler10,Wanderman10,Hopkins06}.

All of the current researches about GRB rate adopt some kind of estimations for 
the survey sensitivities of the instruments for GRB observations.  
A single detection threshold is the most commonly used approximation for estimating the survey sensitivity.
This is generally a good assumption for GRB instruments prior to {\it Swift}.
However, unlike previously flown GRB instruments,
the {\it Swift} Burst Alert Telescope (BAT) adopts a much more complex trigger algorithm in order to
maximize the number of GRB detections.

BAT has over 500 different rate trigger criteria based on photon count rates. 
Each rate trigger criterion adopts a different 
signal-to-noise ratio threshold, different foreground/background time periods for calculating the signal-to-noise ratio,
and covers different energy band \cite{Fenimore03, Graziani03}. 
After an event is triggered by one of the rate-trigger criteria, an image will be generated for further confirmation and localization.
During this imaging process, an additional signal-to-noise ratio based on the real image will be calculated. A rate-triggered event will be confirmed 
as a real detection if (1) the image signal-to-noise ratio is higher than the image threshold (signal-to-noise ratio $\gtrsim 7.0$), and (2) the event is compared with current on-board sky catalog and no known
source is found to be at the event location.

Additional to the rate trigger process, a burst could also be found by an independent ``image trigger'' process,
in which regular image is generated by BAT every minute or longer to search for dim GRBs that are missed by 
the rate trigger.

In order to search for a more robust cosmic GRB rate and study possible systematic effect due to the complex trigger algorithm of BAT, 
we developed a code that is capable of creating mock observed GRB light curves based on adjustable bursts properties, 
and simulating the BAT trigger algorithm for the first time,
including simulating hundreds of rate trigger criteria and mimicking the image threshold. 
We will use this ``BAT-trigger simulator'' to search for a cosmic GRB rate and luminosity
distribution that generate a mock-triggered sample that 
can reproduce the observational GRB characteristics.



\vspace{-10pt}

\section{Comparing with the BAT's sensitivity}
\label{sect:compare_sensitivity}

To test whether our program correctly simulates the complex BAT-trigger algorithm, 
we compare the GRB peak fluxes of the ``triggered'' bursts in our simulation 
to those measured from the real GRBs detected by BAT.

\begin{figure}[!h]
\begin{center}
\includegraphics[width=0.49\textwidth]{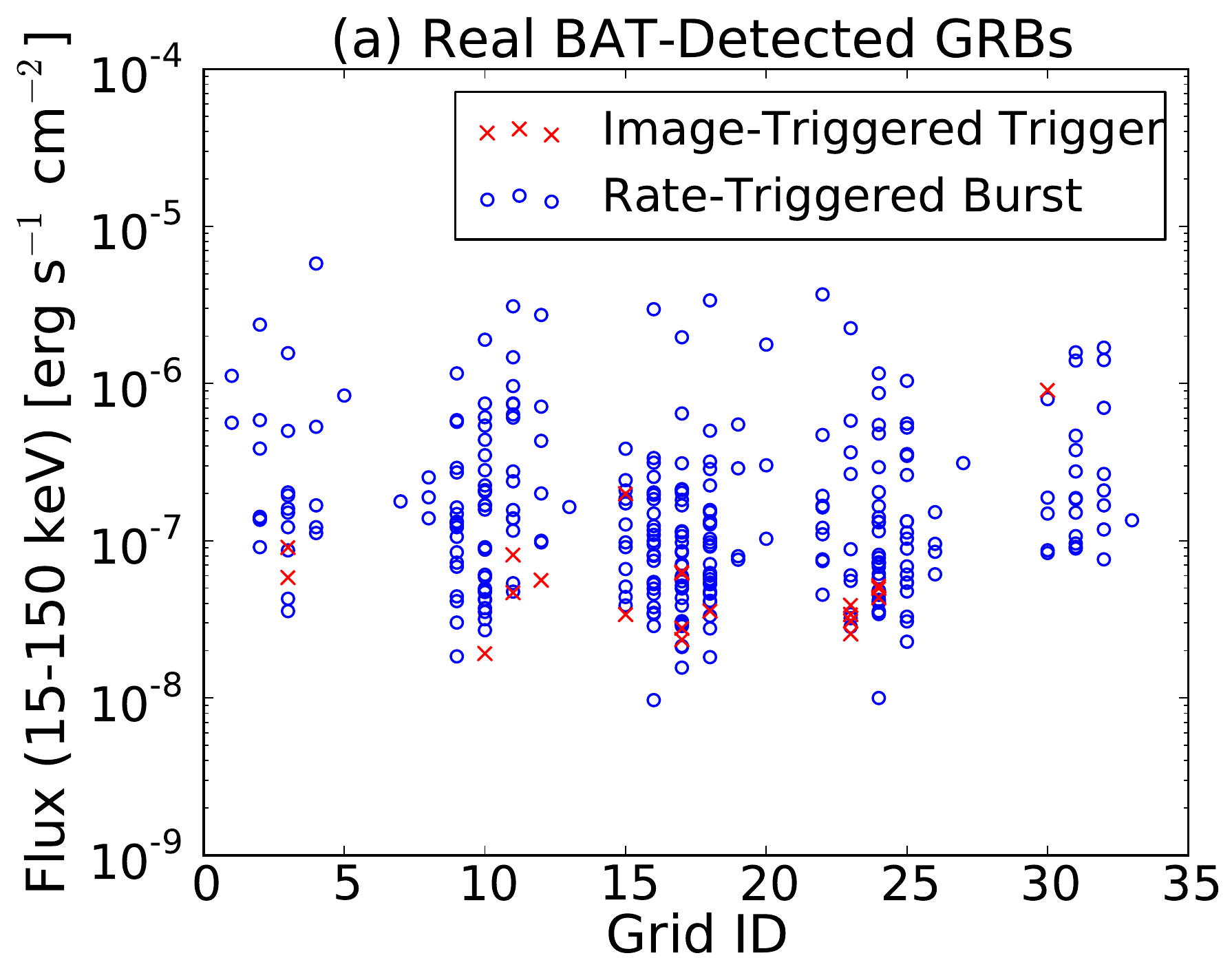}
\includegraphics[width=0.49\textwidth]{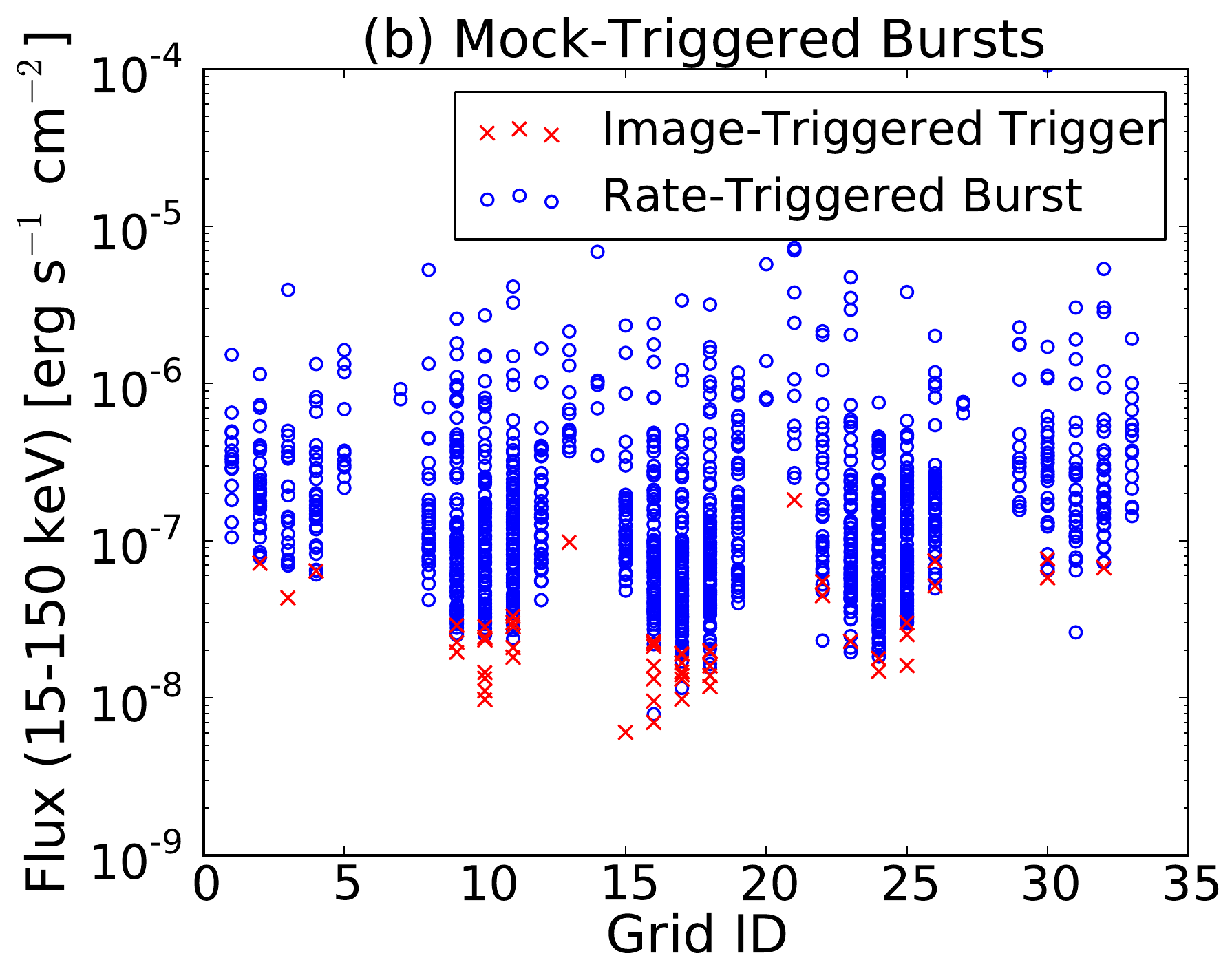}
\end{center}
\caption{
{\it Panel (a)}: 1-s peak energy flux vs. grid ID for the BAT detected bursts from \cite{Sakamoto11}. 
{\it Panel (b)}: Peak energy flux vs. grid ID for the triggered GRBs in the mock sample, assuming 26884 active detectors.
}
\label{fig:sensitivity_grid}
\end{figure}

Fig.~\ref{fig:sensitivity_grid} shows peak fluxes of the real BAT-detected GRBs (panel a) and 
the simulated bursts (panel b), with respect to 
the grid ID of the detector plane.
The grid IDs are simply the number labels on the detector plane, and thus correspond to 
the incoming angles of the bursts relative to the normal axis of the detector.
Each point in the figure represents one burst.
Blue dots in the plot indicate GRBs detected by rate triggers, while red crosses represent bursts found by image triggers.
Results show that our trigger simulator can detect 
GRBs with fluxes $\sim 10^{-8} \ \rm erg \ s^{-1} \ cm^{-2}$ for directly on-axis bursts, and fluxes $\sim 10^{-7} \ \rm erg \ s^{-1} \ cm^{-2}$ for extremely off-axis events,
which is very similar to the real BAT sensitivity.

\vspace{-15pt}

\section{Results of the best-fit parameters and comparison with the star-formation rate}
\label{sect:best_result}

\begin{SCfigure}
\vspace{-15pt}
\includegraphics[width=0.55\textwidth]{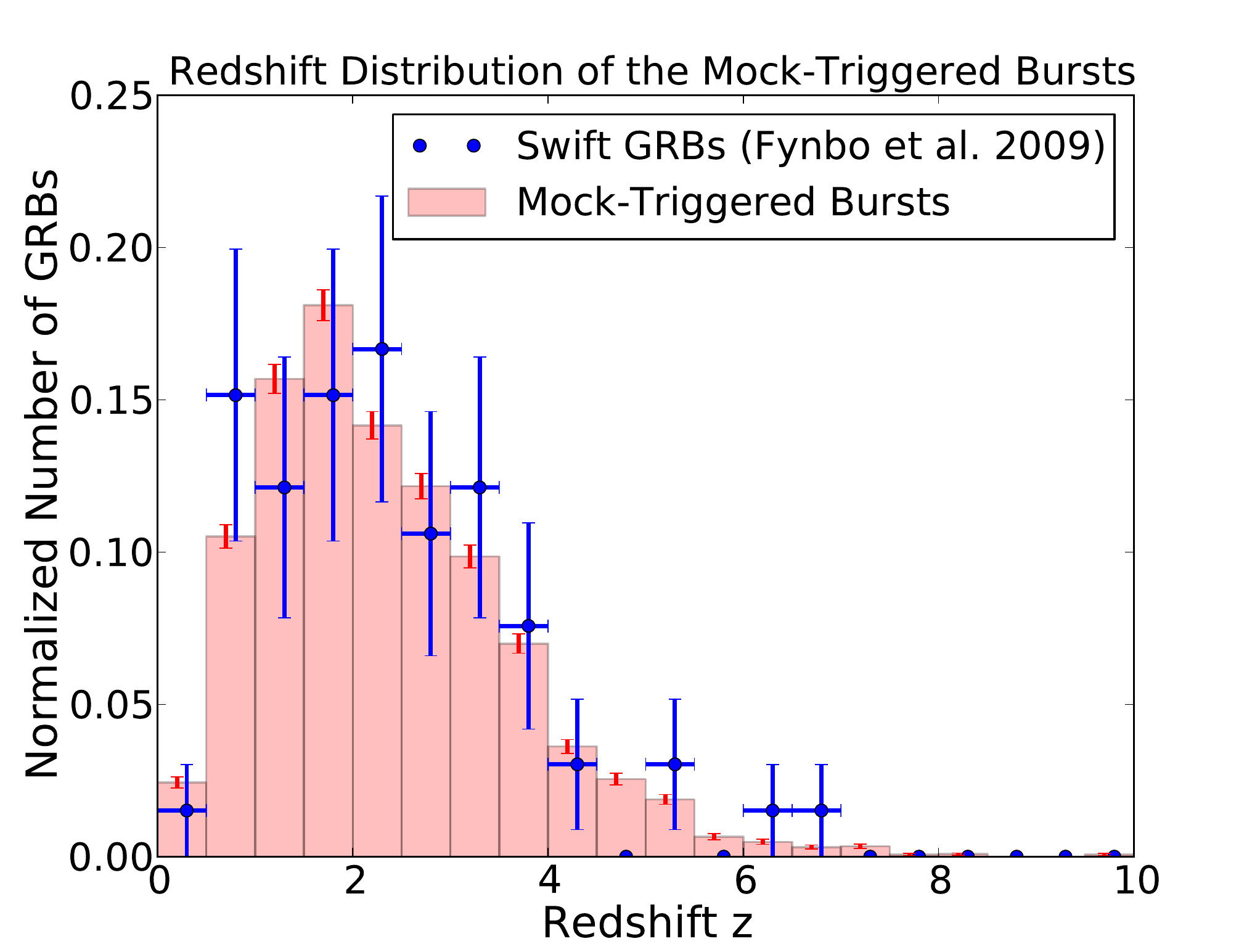}
\vspace{-10pt}
\caption{
The redshift distribution for the mock-triggered bursts (red bars), which are those bursts that are triggered by our trigger simulator.
The redshift distribution of the real BAT-detected bursts are also plotted for comparison (blue dots; \cite{Fynbo09}).
Error bars in the y-axis show the statistical errors in each bin. Error bars in the x-axis represent the bin size. 
}
\label{fig:redshift}
\vspace{+5pt}
\end{SCfigure}

We adopt the functional forms of the GRB rate and luminosity functions in \cite{Wanderman10},
and modify the input parameters in those functions until the mock-triggered sample matches well with observations,
particularly for the redshift and peak-flux distributions.
Figure~\ref{fig:redshift} shows the comparison of the redshift distributions between the mock-triggered bursts 
and real observations.

\begin{wrapfigure}{r}{0.55\textwidth}
\vspace{-30pt}
\begin{center}
\includegraphics[width=0.55\textwidth]{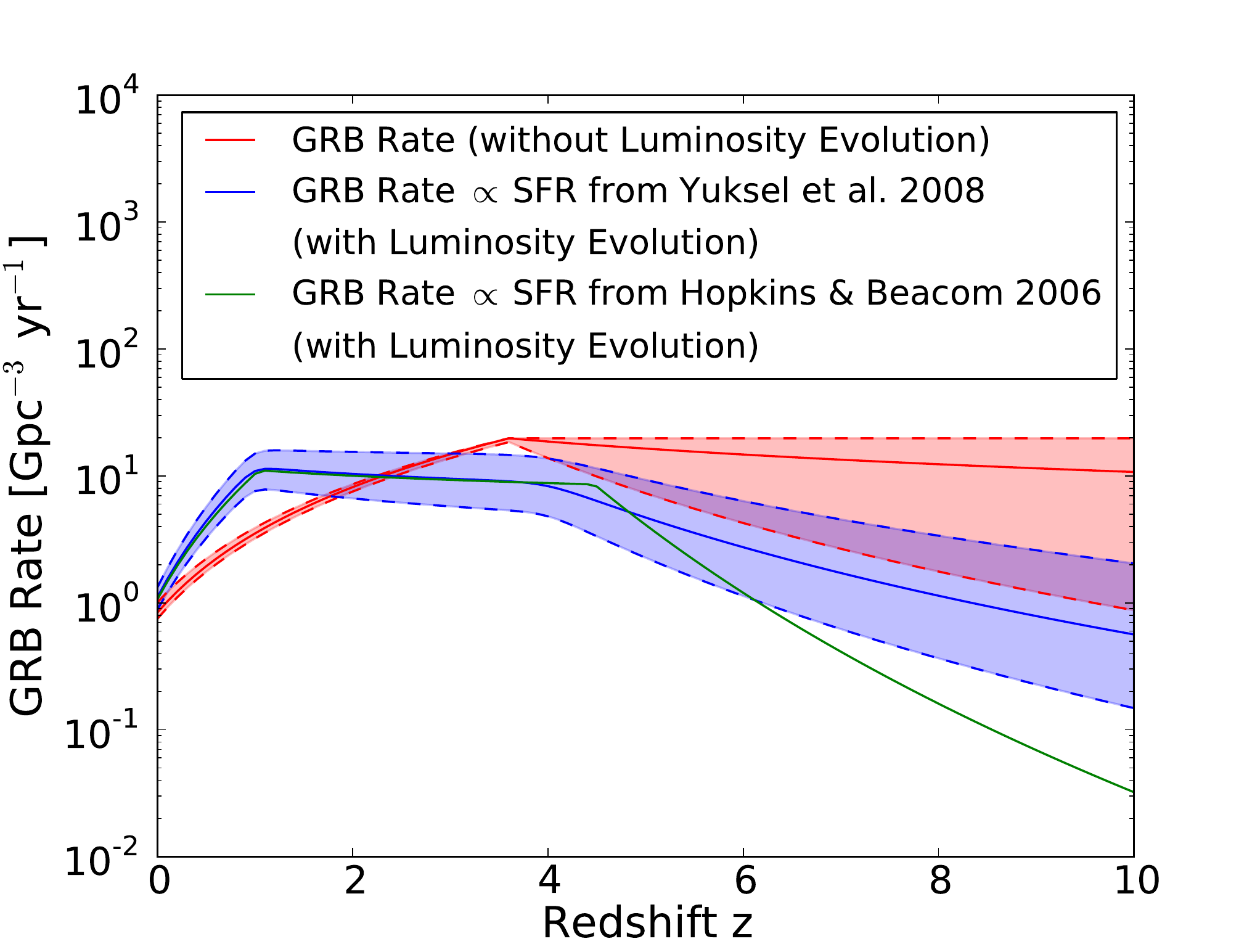}
\end{center}
\vspace{-20pt}
\caption{
Comparison between the cosmic GRB rate from our best-fit sample (red line)
and the cosmic GRB rates that follow strictly the shapes of the SFR from
\cite{Hopkins06} (green line) and \cite{Yuksel08} (blue line).
The GRB rates that trace the SFRs can generate results match well with observational data
if luminosity evolution is included.
}
\label{fig:snr-grb}
\vspace{-40pt}
\end{wrapfigure}

The comparison between our best-fit cosmic GRB rate and the shapes of the cosmic SFR
from \cite{Hopkins06} and \cite{Yuksel08} can be found in Fig.~\ref{fig:snr-grb}.
The SFR from \cite{Yuksel08} is based on the SFR in \cite{Hopkins06}, but 
with correction at high redshift inferred from GRB detections.
Our best-fit GRB rate is plotted in red line.
The green and blue lines in the figure show the GRB rates that strictly follow
the shapes of the SFR from \cite{Hopkins06} and \cite{Yuksel08}, respectively.
The normalization of the green and blue lines come from 
fitting with real GRB observations by including luminosity evolution.

The red and blue shaded regions in Fig.~\ref{fig:snr-grb} show the uncertainties of 
our best-fit GRB rate and the SFR from \cite{Yuksel08}, respectively. For our best-fit GRB rate,
the uncertainty is quantified by modifying the parameters 
in the GRB rate function around our best-fit set of parameters
until results no longer match well with observations (i.e., with statistical significance level $< 90\%$).
The uncertainty of the SFR
measurements from \cite{Yuksel08} is quantified by \cite{Horuichi09}
by taking into account the scatter of data.
Results in Fig.~\ref{fig:snr-grb} show a clear divergence in the shapes of the GRB rate from the SFRs at 
$z \sim 4$. The SFRs, even the one from \cite{Yuksel08}, decline                                                                                                                                                                                                                                                                                                                                                                                                              
much faster than the GRB rate found in our best-fit sample.


\vspace{-15pt}

\section{Conclusions and discussions}

There are several possible explanations for this high GRB rate at large redshift.
First, this could suggest an even larger SFR at high redshift, 
which implies that most of the star formation activities at high redshift probably come from low-luminosity galaxies,
and thus measurements of the SFR based on galaxy observations might underestimate the 
true rate.
Alternatively, a high GRB rate at early times can also be explained if the fraction of GRB-related supernovae
changes as a function of redshift. That is, if there are more supernovae accompanied by GRBs at high redshift, 
one could get a higher GRB rate without adjusting the SFR.

Another possibilities would be the luminosity evolution. 
We found that by allowing the GRB luminosities to evolve with redshift ($L \propto \rm log_{10}(z)$),
both of the GRB rates shown as blue and green lines in Fig.~\ref{fig:snr-grb} 
(i.e., the GRB rates that follow the shapes of SFR from \cite{Yuksel08} and \cite{Hopkins06}, respectively)
can produce results that matches well with observations. 

\vspace{-13pt}

\Acknowledgements
We are grateful for valuable discussions with Brian Fields,
Brett Hayes, Daniel Kocevski, Judith Racusin, Jon Hakkila, Amir Shahmoradi,
Lorenzo Amati, and Michael Briggs.


\vspace{-15pt}

\begin{thebibliography}{99}



\bibitem{Galama98}
Galama, T.~J., et al.\ 1998, Nature, 395, 670

\bibitem{Heger03}
Heger, A., et al. 2003, ApJ, 591, 288

\bibitem{Yuksel08}
Y\"uksel, H., et al. 2008, ApJ, 683, L5

\bibitem{Butler10}
Butler, N. R., et al. 2010, ApJ, 711, 495

\bibitem{Wanderman10}
Wanderman, D., \& Piran, T. 2010, MNRAS, 406, 1944

\bibitem{Hopkins06}
Hopkins, A. M., \& Beacom, J. F. 2006, ApJ, 651, 142

\bibitem{Fenimore03}
Fenimore, E. E., et al. 2003, in American Institute of Physics Conference Series, Vol. 662, Gamma-Ray Burst and Afterglow Astronomy 2001, 491Ð493

\bibitem{Graziani03}
Graziani, C. 2003, in American Institute of Physics Conference Series, Vol. 662, Gamma-Ray Burst and Afterglow Astronomy 2001, 79Ð81

\bibitem{Sakamoto11}
Sakamoto, T., et al. 2011, ApJS, 195, 2

\bibitem{Fynbo09}
Fynbo, J. P. U., et al. 2009, ApJS, 185, 526

\bibitem{Horuichi09}
Horiuchi, S., et al. 2009, Phys. Rev. D, 79, 083013

\end{thebibliography}
\end{document}